# Beam cavity interaction

*A. Gamp*
Deutsches Elektronen-Synchrotron DESY, Hamburg, Germany

**Abstract**
We begin by giving a description of the rf generator–cavity–beam coupled system in terms of basic quantities. Taking beam loading and cavity detuning into account, expressions for the cavity impedance as seen by the generator and as seen by the beam are derived. Subsequently methods of beam-loading compensation by cavity detuning, rf feedback, and feed-forward are described. Examples of digital rf phase and amplitude control for the special case of superconducting cavities are also given. Finally, a dedicated phase loop for damping synchrotron oscillations is discussed.

## 1. Introduction

In modern particle accelerators rf voltages in an extremely large amplitude and frequency range, from a few hundred volts to hundreds of megavolts and from some kHz to many GHz, are required for particle acceleration and storage.

The rf power, which is needed to satisfy these demands can, for example, be generated by triodes, tetrodes, klystrons, or by semiconductor devices. The cw output power available from some tetrodes which were used at HERAp is 60 kW at 208 MHz and up to 800 kW for the 500 MHz klystrons for the new synchrotron light source PETRA III. The 1.3 GHz klystrons for the Free-Electron Laser FLASH at DESY can deliver up to 10 MW rf peak power during pulses of about 1 ms length. Even higher power levels can be obtained from S- and X-band klystrons during pulse lengths on the µs scale.

Such rf power generators generally deliver rf voltages of only a few kV because their source impedance or their output wave guide impedance is small as compared to the shunt impedance of the cavities in the accelerators.

Typically, a tetrode has its highest efficiency for a load resistance of less than a kΩ whereas the cavity shunt impedance usually is of the order of several MΩ. This is the real impedance, which the cavity represents to a generator at the resonant frequency. It must not be confused with ohmic resistances.

Optimum fixed impedance matching between generator and cavity can easily be achieved with a coupling loop in the cavity. There is, however, the complication that the transformed cavity impedance as seen by the generator depends also on the synchronous phase angle and on the beam current and is therefore not constant as we will show quantitatively. The beam current induces a voltage in the cavity, which may become even larger than the one induced by the generator. Owing to the vector addition of these two voltages the generator now sees a cavity which appears to be detuned and unmatched except for the particular value of beam current for which the coupling has been optimized. The reflected power occuring at all other beam currents has to be handled.

In addition, the beam-induced cavity voltage may cause single- or multi-bunch instabilities, since any bunch in the machine may see an important fraction of the cavity voltage induced by itself or from previous bunches. This voltage is given by the product of beam current and cavity impedance as seen by the beam. Minimizing this latter quantity is therefore essential. It is also called beam loading compensation, and some servocontrol mechanisms, which can be used to achieve this goal, will be discussed in this lecture.

## 2. The coupling between the rf generator, the cavity, and the beam

For frequencies in the neighbourhood of the fundamental resonance an rf cavity can be described [1] by an equivalent circuit consisting of an inductance $L_2$, a capacitor $C$, and a shunt impedance $R_S$ as shown in Fig. 1. In practice, $L_2$ is made up by the cavity walls whereas the coupling loop $L_1$ is usually small as compared to the cavity dimensions.

In this example a triode with maximum efficiency for a real load impedance $R_A$ has been taken as an rf power generator. For simplicity we consider a short and lossless transmission line between the generator and $L_1$. Then there is optimum coupling between the generator and the empty (i.e, without beam) cavity for

$$N^2 = R_S / R_A = L_2 / L_1 , \tag{1}$$

where, for maximum power output, $R_A$ equals the dynamic source impedance $R_I$. $N$ is called the transformation or step-up ratio.

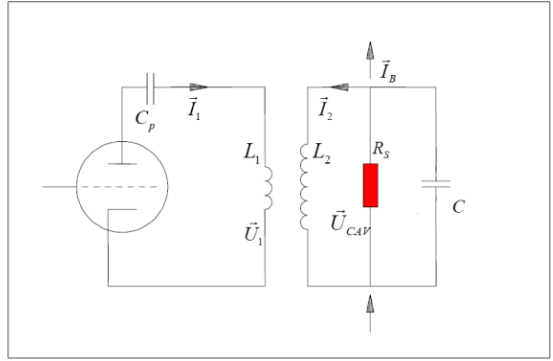

**Fig. 1:** Equivalent circuit of a resonant cavity near its fundamental resonance. In practice, the inductance $L_2$ is made up by the cavity walls, whereas $L_1$ usually is a small coupling loop. The capacitor $C$ denotes the equivalent cavity capacitance whereas $C_p$ is needed only for separation of the plate dc voltage from the rest of the circuit.

Since, in general, there may be power transmitted from the generator to the cavity and also, in the case of imperfect matching, vice versa, the voltage $\vec{U}_1$ is expressed as the sum of two voltages

$$\vec{U}_1 = \vec{U}_{forward} + \vec{U}_{reflected} , \tag{2}$$

whereas the corresponding currents flow in the opposite directions, hence

$$\vec{I}_1 = \vec{I}_{forward} - \vec{I}_{reflected} . \tag{3}$$

The minus sign in Eq. (3) indicates the counterflowing currents while voltages of forward and backward waves just add up. So, in the simplest case where the beam current $\vec{I}_B = 0$ and where the generator frequency $f_{Gen} = f_{Cav}$, there is no reflected power from the cavity to the generator, and $\vec{U}_1$ and $\vec{I}_1$ are identical to the generator voltage and current, respectively. One has

$$\vec{U}_{CAV} = N\vec{U}_1 . \tag{4}$$

Now we can derive an expression for the complex cavity voltage as a function of generator and beam current and of the cavity and generator frequency.

According to Fig. 1 the cavity voltage $\vec{U}_{CAV}$ can be written as

$$\vec{U}_{CAV} = L_2 \left( \dot{\vec{I}}_2 + \dot{\vec{I}}_1 / N \right) \tag{5}$$

$$\vec{I}_2 = -\left( \vec{I}_B + \vec{U}_{CAV} / R_S + C \dot{\vec{U}}_{CAV} \right). \tag{6}$$

All voltages and currents have the time dependence

$$\vec{U} = \hat{\vec{U}} e^{i\omega t}. \tag{7}$$

$\vec{I}_B = \vec{I}_B(\omega)$ is the harmonic content at the frequency $\omega$ of the total beam current. Throughout this article we consider only a bunched beam with a bunch spacing small as compared to the cavity filling time. In this case $\vec{I}_B(\omega)$ is quasi sinusoidal. We also restrict the discussion to the interaction of the beam with the fundamental cavity resonance. The interaction with higher order cavity modes can be minimized by dedicated damping antennas built into the cavity.

Inserting Eq. (6) in Eq. (5) and using

$$2\pi f_{CAV} = \omega_{CAV} = \frac{1}{\sqrt{L_2 C}}, \tag{8}$$

one finds

$$\omega_{CAV}^2 \vec{U}_{CAV} = \frac{1}{C} \left[ \frac{1}{N} \dot{\vec{I}}_1 - \dot{\vec{I}}_B - \frac{1}{R_S} \dot{\vec{U}}_{CAV} \right] - \ddot{\vec{U}}_{CAV}. \tag{9}$$

We define

$$\Gamma = \frac{1}{2CR_S} = \frac{\omega_{CAV}}{2Q} \tag{10}$$

where the quality factor of the cavity can be expressed as $2\pi$ times the ratio of total electromagnetic energy stored in the cavity to the energy loss per cycle.

Here we would like to mention that the ratio

$$\frac{R_S}{Q} = \sqrt{\frac{L_2}{C}} \tag{11}$$

is a characteristic quantity of a cavity depending only on its geometry. We can rewrite Eq. (9) as

$$\ddot{\vec{U}}_{CAV} + 2\Gamma \dot{\vec{U}}_{CAV} + \omega_{CAV}^2 \vec{U}_{CAV} = 2\Gamma R_S \left[ \frac{1}{N} \dot{\vec{I}}_1 - \dot{\vec{I}}_B \right]. \tag{12}$$

This equation describes a resonant circuit excited by the current $\vec{I} = (\vec{I}_1 / N - \vec{I}_B)$. The minus sign occurs because the generator-induced cavity voltage has opposite sign to the beam-induced voltage, which would decelerate the beam. It can be shown that the beam actually sees only 50% of its own induced voltage. This is called the fundamental theorem of beam loading [2, 3].

### 2.1 The impedance of the generator-loaded cavity as seen by the beam

In order to find the cavity impedance as seen by the beam, we make use of Eqs. (2), (3), and (4) to express the generator current term of Eq. (12) in the form

$$\frac{1}{N}\dot{\vec{I}}_1 = \frac{1}{NR_A}\left[2\dot{\vec{U}}_{forward} - \dot{\vec{U}}_1\right] = \frac{1}{N}\left[2\dot{\vec{I}}_{forward} - \frac{\dot{\vec{U}}_{CAV}}{NR_A}\right]. \tag{13}$$

The new term $\dfrac{\dot{\vec{U}}_{CAV}}{NR_A}$ leads to a modification of the damping term in Eq. (12):

$$\ddot{\vec{U}}_{CAV} + 2\Gamma(1+\beta)\dot{\vec{U}}_{CAV} + \omega_{CAV}^2 \vec{U}_{CAV} = 2\Gamma_L R_{SL}\left[\frac{2}{N}\dot{\vec{I}}_f - \dot{\vec{I}}_B\right]. \tag{14}$$

With the coupling ratio

$$\beta = R_S/(N^2 R_A) \tag{15}$$

we can introduce the 'loaded' damping term

$$\Gamma_L = \Gamma(1+\beta) \tag{16}$$

and consequently, in accordance with Eq. (10), the loaded cavity $Q$ and loaded shunt impedance are

$$Q_L = Q/(1+\beta) \quad \text{and} \quad R_{SL} = R_S/(1+\beta). \tag{17}$$

In the case of perfect matching in the absence of beam, i.e., $\beta = 1$, the damping term simply doubles and $Q$ and $R_S$ take half their original values. This is due to the fact that the beam would see the cavity shunt impedance $R_S$ in parallel or loaded with the transformed generator impedance $N^2 R_A = R_S$. Therefore we find in Eq. (14) that the transformed generator current

$$\vec{I}_G = 2\vec{I}_f / N \tag{18}$$

gives rise to twice as much cavity voltage as a similar beam current would do. Here and in Eq. (15) we assume that the transformed dynamic source impedance $N^2 R_A$ is identical to generator impedance seen by the cavity. This is strictly true only if a circulator is placed between the rf power generator and the cavity. Without a circulator it may be approximately true if the power source is a triode. Owing to its almost constant anode voltage to current characteristic the impedance of a tetrode as seen from the cavity is, however, much bigger than the corresponding $R_A$ and therefore $R_{SL} \approx R_S$ in this case where a short transmission line (or of length $n\lambda/2, n$ integer) is considered.

Following Ref. [4] we write the solution of Eq. (14) in the Fourier–Laplace representation

$$\hat{\vec{U}}_{CAV} = \frac{i\omega}{\omega_{CAV}^2 - \omega^2 + 2i\omega\Gamma_L} 2\Gamma_L R_{SL}\left[\hat{\vec{I}}_G - \hat{\vec{I}}_B\right]. \tag{19}$$

For $\Delta\omega \ll \omega_{CAV}$ this can be approximated by

$$\hat{\vec{U}}_{CAV} \approx \frac{R_{SL}\left[\hat{\vec{I}}_G - \hat{\vec{I}}_B\right]}{1 + iQ_L 2\dfrac{\Delta\omega}{\omega_{CAV}}} \tag{20}$$

where $\omega = \omega_{CAV} + \Delta\omega$.

A plot of the cavity voltage modulus and its real and imaginary part as a function of $\Delta\omega$ is shown in Fig. 5. For a resonant cavity the beam-induced voltage $\hat{\vec{U}}_B$, or the beam loading, is thus given by the product of loaded shunt impedance and beam current:

$$\hat{\vec{U}}_B = -R_{SL}\hat{\vec{I}}_B. \tag{21}$$

The ideal beam loading compensation would, therefore, minimize $R_{SL}$ without increasing the generator power necessary to maintain the cavity voltage.

The beam-induced voltages are by no means negligible. For a loaded shunt impedance of, say 2.5 MΩ, and a beam current of 0.2 A, the induced voltage would be 0.5 MV! To compensate this, a generator current of 20 A would be needed for a typical transformation ratio $N = 100$. This may lead to large values of reflected power which must be taken into consideration when designing the rf system.

## 2.2 The impedance of the beam-loaded cavity as seen by the generator

Having just discussed the impedance, which the combined system generator + cavity represents to the beam we would like to discuss in the following the impedance $Z$, or rather admittance $Y = 1/Z$, which the combined cavity and beam system represents to the generator.

From Eqs. (1), (5) and (6) one sees [5] that

$$Y = \frac{\vec{I}_1}{\vec{U}_1} = \frac{N^2}{R_S} + \frac{\vec{I}_B N^2}{\vec{U}_{CAV}} + \frac{N^2}{i\omega L_2}\left(1 - \frac{\omega^2}{\omega_{CAV}^2}\right) \tag{22}$$

which reduces to $Y = N^2/R_S = 1/R_A$ for a tuned cavity without beam current in the case of $\beta = 1$. As we are now going to show, a non-vanishing real part of the quotient $\vec{I}_B/\vec{U}_{CAV}$ will necessitate a change in $\beta$ to maintain optimum matching whereas the imaginary part can be compensated by detuning the cavity. In order to work out Re and $\text{Im}(\vec{I}_B/\vec{U}_{CAV})$ we define the angle $\phi_s$ as the phase angle between the synchronous particle and the zero crossing of the rf cavity voltage. The accelerating voltage is therefore given by

$$\vec{U}_{ACC} = \vec{U}_{CAV}\sin\phi_s \tag{23}$$

and the normalized cavity voltage and beam current are related by

$$\frac{\vec{I}_B}{\vec{U}_{CAV}} = \frac{|\vec{I}_B|}{|\vec{U}_{CAV}|}e^{i\left(\frac{\pi}{2}-\phi_s\right)}. \tag{24}$$

Consequently

$$\text{Re}\left(\frac{\vec{I}_B}{\vec{U}_{CAV}}\right) = \left|\frac{\vec{I}_B}{\vec{U}_{CAV}}\right|\sin\phi_s \tag{25}$$

and

$$\text{Im}\left(\frac{\vec{I}_B}{\vec{U}_{CAV}}\right) = \left|\frac{\vec{I}_B}{\vec{U}_{CAV}}\right|\cos\phi_s. \tag{26}$$

The real part of the admittance seen by the generator then becomes

$$\text{Re}(Y) = \frac{N^2}{R_S}\left(1 + \frac{R_S |\vec{I}_B|}{|\vec{U}_{CAV}|}\sin\phi_s\right). \qquad (27)$$

We see that the term in the bracket describes a change in admittance caused by the beam. In order to maintain optimum coupling the coupling ratio $\beta$ must now take the value

$$\beta = \left(1 + \frac{R_S |\vec{I}_B|}{|\vec{U}_{CAV}|}\sin\phi_s\right). \qquad (28)$$

This result tells us that the change in the real part of the admittance is proportional to the ratio of rf power delivered to the beam to rf power dissipated in the cavity walls. For circular electron machines, where the considerable amount of energy lost by synchrotron radiation has to be compensated continuously by rf power, values of $\phi_s \geq 30^0$ and $\beta \geq 1.2$ are typical for high beam current and normal conducting cavities.

A typical set of parameters for this case would be $R_S = 6\,\text{M}\Omega, \vec{U}_{CAV} = 1\,\text{MV}$ and $\vec{I}_B(\omega) = 60\,\text{mA}$. This implies, of course, that for a $\beta$, which has been optimized for the maximum beam current, there will be reflected generator power for lower beam intensities. If the power source is a klystron, this can be handled by inserting a circulator in the path between generator and cavity or, in the case of a tube, by a sufficiently high plate dissipation power capability.

For superconducting cavities the situation is totally different. Here a typical set of parameters would be $R_S = 10^{13}\,\Omega, \vec{U}_{CAV} = 25\,\text{MV}$, and $\vec{I}_B(\omega) = 16\,\text{mA}$ and $\varphi_s = 90°$. Then $\beta = 6401$, and for a typical unloaded $Q = 10^{10}$ the loaded quantity becomes $Q_L = 1.6*10^6$. If the loaded $Q$ is adjusted to this value, so that there is no reflection of rf power back to the cavity at the nominal beam current, it means also that there is a strong mismatch and hence almost total reflection without beam.

The complex reflection coefficient is given by

$$r(\beta,\Delta\omega) = \frac{\beta - 1 - \dfrac{iQ2\Delta\omega}{\omega_{cav}}}{\beta + 1 + \dfrac{iQ2\Delta\omega}{\omega_{cav}}}. \qquad (29)$$

On resonance it simplifies to

$$r(\beta) = \frac{\beta - 1}{\beta + 1} = \frac{Z - Z_0}{Z + Z_0} = \frac{\vec{U}_{reflected}}{\vec{U}_{forward}} = \frac{6400}{6402} \approx 1. \qquad (30)$$

The voltage standing wave ratio (VSWR) then becomes

$$VSWR = \frac{\vec{U}_{forward} + \vec{U}_{reflected}}{\vec{U}_{forward} - \vec{U}_{reflected}} = \frac{1 + r}{1 - r} \approx \infty. \qquad (31)$$

So, for superconducting cavities, beam loading is even more dramatic than it may be for normal conducting cavities since situations where total reflection of the incident generator power occurs during significant time intervals are unavoidable.

It is instructive to look at the time dependence of the envelope of the cavity voltage and of the reflected voltage. The solution of Eq. (14) yields for the envelope of the cavity voltage during filling

$$\vec{U}_{CAV}(t) \approx \frac{\hat{\vec{U}}_{CAV}(1-e^{-(1/\tau - i\Delta\omega)t})}{1+iQ_L 2\frac{\Delta\omega}{\omega_{CAV}}} \qquad (32)$$

with the time constant

$$\tau = 2Q_L/\omega_{CAV} . \qquad (33)$$

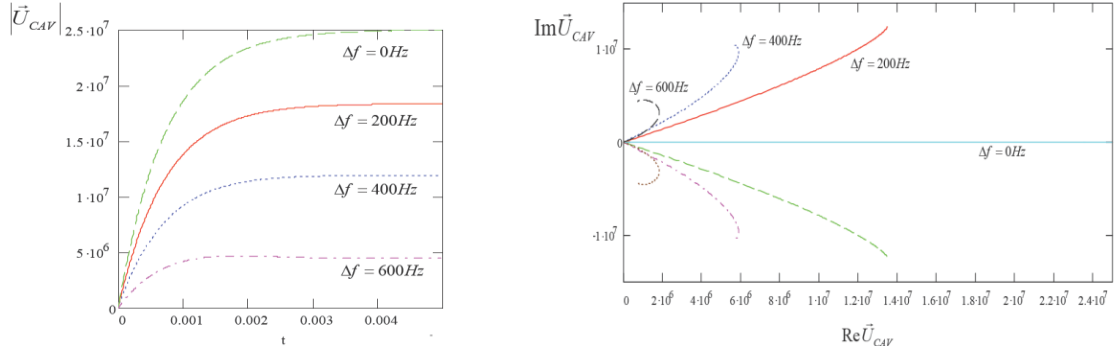

**Fig. 2a and b:** Modulus and real and imaginary part of the cavity voltage calculated with Eq. (32) as a function of detuning frequency

On resonance, Eq. (32) simplifies to

$$\vec{U}_{CAV}(t) = \hat{\vec{U}}_{CAV}(1-e^{-t/\tau}) . \qquad (34)$$

From the value $P_{CAV}$ of the power transmitted into the cavity

$$P_{CAV} = P_{forward} - P_{reflected} = P_{forward}(1-r^2) = P_{forward}\frac{4\beta}{(1+\beta)^2} \qquad (35)$$

the asymptotic value $\vec{U}_{CAV}(\infty) = \hat{\vec{U}}_{CAV}$ can be obtained as a function of $\beta$:

$$\hat{\vec{U}}_{CAV} = \sqrt{2R_S P_{forward}\frac{4\beta}{(1+\beta)^2}} . \qquad (36)$$

The reflected voltage can be expressed in terms of the forward voltage and β:

$$\vec{U}_{reflected} = \vec{U}_{forward}\frac{\beta-1}{\beta+1} \qquad \vec{U}_1 = \vec{U}_{forward} + \vec{U}_{reflected} = \vec{U}_{forward}\frac{2\beta}{1+\beta} . \qquad (37)$$

From Eqs. (37) and (2) one finds

$$\vec{U}_{reflected}(t) = \frac{1}{N}\hat{\vec{U}}_{CAV}(1-e^{-t/\tau}) - \vec{U}_{forward} = \hat{\vec{U}}_1\left[(1-e^{-t/\tau}) - \frac{1+\beta}{2\beta}\right]. \qquad (38)$$

For the matched case where $\beta = 1$ one sees that the reflected voltage reaches $0$ asymptotically as the cavity voltage reaches the value $\vec{U}_{CAV}(\infty) = \hat{\vec{U}}_{CAV}$ given by Eq. (36). For $\beta \gg 1$, however, $\vec{U}_{reflected}(t) = 0$ only at the time

$$t_{U_{refl}=0} = -\tau \ln(1 - \frac{1+\beta}{2\beta}). \tag{39}$$

At this time the cavity voltage has reached exactly the voltage for which $\beta$ has been calculated with Eq. (28) for a given beam current, which is about half of the asymptotic value:

$$\vec{U}_{CAV}(t_{U_{refl}=0}) = \hat{\vec{U}}_{CAV} \frac{1+\beta}{2\beta} \approx 0.5 \hat{\vec{U}}_{CAV}. \tag{40}$$

This can be illustrated by taking the beam-induced voltage into account when calculating the envelope of the cavity voltage. In Eq. (41) the case where the beam is injected at $t_{U_{refl}=0}$ is considered:

$$\vec{U}_{CAV}(t) = \hat{\vec{U}}_{CAV}(1-e^{-t/\tau}) - \hat{\vec{U}}_{Beam}(1-e^{-t(>t_{U_{refl}=0})/\tau}). \tag{41}$$

This is shown in Fig. 4. The beam-induced voltage increases with the same time constant as the cavity voltage, but starting only at $t_{U_{refl}=0}$ and, in this example, with opposite sign. Therefore the sum of the two voltages stays constant for $t > t_{U_{refl}=0}$.

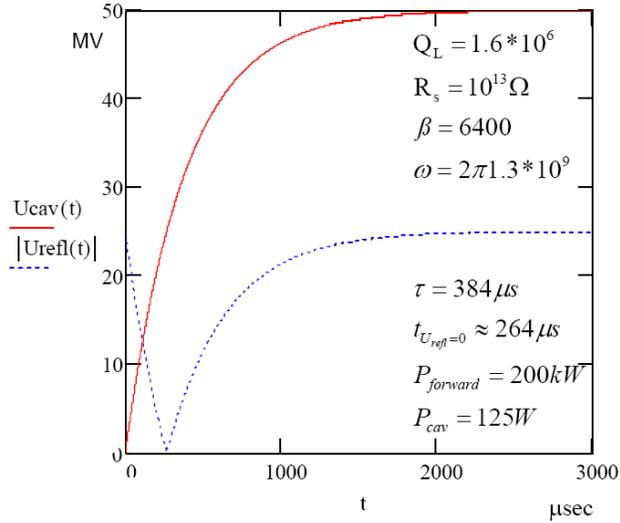

**Fig. 3:** Plot of the envelope of the cavity voltage and modulus of the reflected voltage calculated from Eqs. (34) and (38)

So far we have seen that pure real beam loading, where $\hat{\vec{U}}_B$ and $\hat{\vec{U}}_{Gen}$ are either in phase or opposite, can be compensated by adjustment of $\beta$ and generator power. Now we will show that in contrast to this, pure reactive beam loading, where $\hat{\vec{U}}_B$ and $\hat{\vec{U}}_{Gen}$ are in quadrature, can be compensated by detuning the cavity. This means that the original cavity voltage can be restored by detuning the cavity. No additional generator power is needed in steady state, but for transient beam loading compensation significantly more power may be needed.

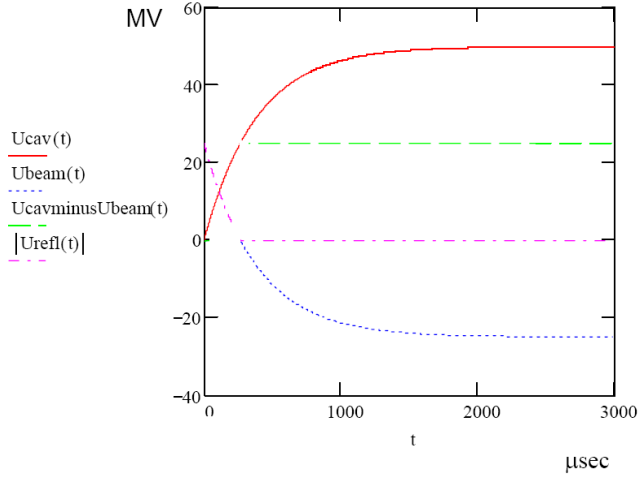

**Fig. 4:** Same as Fig. 3, but with the beam injected at $t_{U_{refl}=0}$. Then, for $t > t_{U_{refl}=0}$ the reflected voltage stays 0, since now there is matching with beam, and the cavity voltage stays constant since both generator-induced and beam-induced voltages increase with the same time constant but with opposite sign. This is indicated by the dashed line calculated with Eq. (41) which coincides with the full line for $t < t_{U_{refl}=0}$.

From the imaginary part of Eq. (22) and from Eq. (26) we find that the apparent cavity detuning caused by the beam current can be compensated by a real cavity detuning (for example, by means of a mechanical plunger cavity tuner) of the amount

$$\frac{\omega}{\omega_{CAV}} = \sqrt{1 + \frac{R_S |\vec{I}_B|}{Q |\vec{U}_{CAV}|} \cos\phi_s} \ . \qquad (42)$$

Expanding the square root to first order we find a cavity detuning angle $\Psi$

$$\tan\Psi \approx \frac{R_{SL} |\vec{I}_B|}{|\vec{U}_{CAV}|} \cos\phi_s \approx 2Q_L \frac{\Delta\omega}{\omega_{CAV}} \ . \qquad (43)$$

This is essentially the ratio between beam-induced and total cavity voltage.

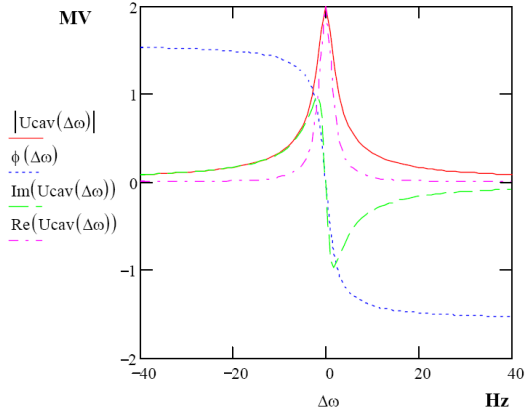

**Fig. 5:** Plot of the envelope of the cavity voltage modulus, its real and imaginary part calculated with Eq. (20) and of the detuning angle calculated with Eq. (43)

In order to calculate the maximum amount of reflected power seen by the generator as a consequence of beam-loading we consider, for $\beta = 1$, a tuned cavity, i.e., $\omega = \omega_{CAV}$. Then, with Eqs. (25) and (26), Eq. (22) reads

$$Y = \frac{1}{R_A}\left[1 + \frac{R_S|\vec{I}_B|}{|\vec{U}_{CAV}|}\sin\phi_S + i\frac{R_S|\vec{I}_B|}{|\vec{U}_{CAV}|}\cos\phi_S\right]. \tag{44}$$

Solving for $\vec{U}_{refl.}$ by means of Eq. (2) and Eq. (3) the reflected power $P_{refl.} = \left|\hat{\vec{U}}_{refl.}\right|^2/2R_A$ becomes

$$P_{refl.} = R_S \hat{\vec{I}}_B^2 / 8. \tag{45}$$

This corresponds to half of the power given by the beam to the coupled system cavity + generator. The second half of this power is dissipated in the cavity walls. All we found is that two equal resistors in parallel dissipate equal amounts of power. As we pointed out above, this is strictly true only if a circulator is placed in between the rf power source and the cavity. Nevertheless, the amount of reflected power can be quite impressive. For an average dc beam current of, say, 0.1 A the harmonic current $\hat{\vec{I}}_B(\omega)$ may become up to twice as large. Then, taking $R_S = 8\,\text{M}\Omega$, for example, we find 40 kW of reflected power, which have to be dissipated.

For a cavity where only the reactive part of the beam-loading has been compensated by detuning according to Eq. (43), but $\beta = 1$, the reflected power is given by

$$P_{refl.} = R_S \hat{\vec{I}}_B^2 \sin^2\phi_s / 8. \tag{46}$$

Summarizing the results of this section we state that the beam sees the cavity shunt impedance in parallel with the transformed generator impedance. The resulting loaded impedance is reduced by the factor $1/(1+\beta)$. The optimum coupling ratio between generator and cavity depends on the amount of energy taken by the beam out of the rf field. The coupling is usually fixed and optimized for maximum beam current. The amount of cavity detuning necessary for optimum matching, on the other hand, depends on the ratio of beam-induced to total cavity voltage. Clearly these issues depend also on the synchronous phase angle.

### 3. Beam-loading compensation by detuning

In Fig. 8 a diagram of a tuner regulation circuit is shown. The phase detector measures the relative phase between generator current and cavity voltage which depends, according to Eq. (43), on the frequency $\Delta\omega$ by which the cavity is detuned. The phase detector output signal acts on a motor which drives a plunger tuner into the cavity volume until there is resonance. An alternative tuner could be a resonant circuit loaded with ferrites. The magnetic permeability $\mu$ of the ferrites and hence the resonance frequency of the circuit can be controlled by a magnetic field. This latter method is especially useful when a large tuning range in combination with a low cavity $Q$ is required.

If proper tuner action is necessary in a large dynamic range of cavity voltages, limiters with a minimum phase shift per dB compression have to be installed at the phase detector input. Since this phase shift decreases with frequency, all signals should be mixed down to a sufficiently low intermediate frequency.

The signal proportional to the generator current $\vec{I}_{forw.}$ can be obtained from a directional coupler. In case the rf amplifier is so closely coupled to the cavity that no directional coupler can be installed, the relative phase between rf amplifier input and output signal can also be used to derive a tuner signal [6].

As we have shown in the previous section, stationary beam-loading can be entirely compensated by detuning the cavity, provided that the synchronous phase angle is small or zero. This is usually the case in proton synchrotrons during storage, where the energy loss due to the emission of synchrotron radiation is negligible. Here, the rf voltage is needed only to keep the bunch length short. Energy ramping also takes place at very small $\phi_s$.

In the following, we will restrict ourselves, for simplicity, to hadron machines. Consequently $\beta = 1$, $\phi_s \approx 0$, and the generator- and beam-induced voltages are in quadrature. There are, however, also in this case, several limitations to detuning as the only means of beam-loading compensation. One is known as Robinson's stability criterion [7], which we will briefly explain here.

We consider a perturbation voltage $\vec{U}_{pertub}(t) = \hat{\vec{U}}_{perturb} e^{i\Omega t}$ such that

$$\vec{U}_{CAV}(t) = (\hat{\vec{U}}_{CAV} + \hat{\vec{U}}_{perturb} e^{i\Omega t}) e^{i\omega_{CAV} t}. \tag{47}$$

If $\Omega$ is close to $\Omega_S$, a coherent synchrotron oscillation of all bunches with a damping constant $D_S$ may be excited. This oscillation leads to two new frequency components $\omega \pm \Omega$ in the beam current frequency spectrum. These two components will induce additional rf voltages in the cavity. Their amplitudes are unequal, since $Z_{Cav}(\omega) \approx R_{SL}/(1 + iQ_L 2\Delta\omega/\omega_{CAV})$ and hence, with $R(\omega) = \mathrm{Re}\, Z(\omega)$,

$$R(\omega + \Omega) \neq R(\omega - \Omega). \tag{48}$$

These two induced voltages act back on the beam current, and when the induced voltage has the same phase and larger amplitude than the perturbation voltage the oscillation will grow and become instable. The stability condition can be written as

$$\frac{R(\omega+\Omega) - R(\omega-\Omega)}{\vec{U}_{CAV} \sin\phi_S} \vec{I}(\omega) < 4 \frac{D_S}{\Omega_S} \tag{49}$$

where $\omega = \omega_{beam} = h\omega_{revolution}$ and $h$ = harmonic number.

This result from Piwinski [5], which agrees with the Robinson criterion, is illustrated in Fig. 6.

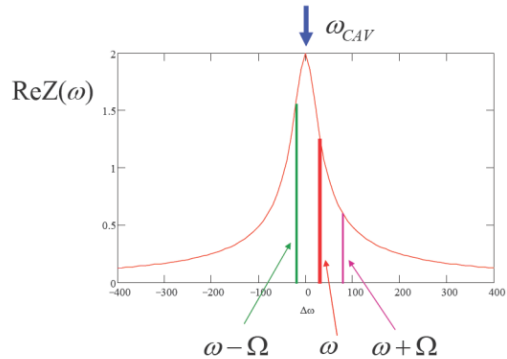

**Fig. 6:** Illustration of a Robinson-stable scenario since $\omega_{CAV} < \omega$ and hence $R(\omega+\Omega) < R(\omega-\Omega)$ (see text)

The situation becomes more complex when there are additional resonances or cavity modes close to other revolution harmonics of the beam current $\vec{I}_B(\omega)$ which may also lead to instabilities. Also the spectrum of the beam can become much more complicated as is schematically indicated in Fig. 7 where only the fundamental synchrotron oscillation mode is drawn.

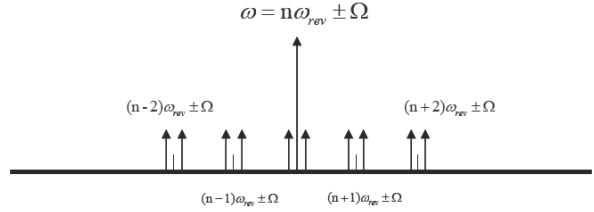

**Fig. 7:** Example of a beam spectrum with nearby revolution harmonics and synchrotron frequency sidebands

Damping of synchrotron oscillations can be achieved by several means. One possibility consists of an additional passive cavity with an appropriate resonance to change $R(\omega+\Omega)$ and $R(\omega-\Omega)$ such that stability criterion (49) is fulfilled.

Another possibility is an additional acceleration voltage with slightly smaller frequency to separate the synchrotron frequencies of different bunches such that the oscillation is damped by decoherence. An active phase loop for damping synchrotron oscillations will be described in the last section.

The beam will also become unstable if the amount of detuning calculated by Eq. (43) becomes comparable to the revolution frequency of the particles in a synchrotron. The finite time of, say, a second, which is needed for the tuner to react, can also create instabilities. Actually, the time scale of the cavity voltage transients, which may cause beam instabilities, is much shorter. According to Eq. (34) the cavity voltage rise after injection of a bunched beam with a current $\vec{I}_B(\omega_{CAV})$ can be approximated by

$$\vec{U}_B \approx R_{SL} \vec{I}_B \left(1 - e^{-t/\tau}\right). \tag{50}$$

This voltage will add to the cavity voltage produced by the generator, and after a time $t \approx 3\tau$ the total cavity voltage becomes

$$\left|\vec{U}_{CAV}\right| \approx R_{SL} \sqrt{\left|\vec{I}_g\right|^2 + \left|\vec{I}_B\right|^2} \tag{51}$$

with a phase shift given by Eq. (43).

Since, for normal conducting cavities, typical values of $\tau$ are below 100 μs and therefore much smaller than the proton synchrotron frequency in a storage ring ($T_S$ is usually $\geq$ some ms), these transients will, in general, excite synchrotron oscillations of the beam with the consequence of emittance blow-up and particle loss or even total beam loss. Additional compensation of transient beam-loading is therefore necessary. Individual phase and amplitude loops may become unstable due to the correlation of both quantities [8, 9].

In the following section we discuss fast feedback as a possibility to overcome these problems.

## 4. Reduction of transient beam-loading by fast feedback

The principle of a fast feedback circuit is illustrated in Fig. 8. A small fraction $\alpha$ of the cavity rf signal is fed back to the rf preamplifier input and combined with the generator signal. The total delay $\delta$ in the feedback-path is such that both signals have opposite phase at the cavity resonance frequency. For other frequencies there is a phase shift

$$\Delta\varphi = \Delta\omega\delta .\tag{52}$$

Therefore the voltage at the amplifier input is now given by

$$\vec{U}'_{in} = \vec{U}_{in} - e^{-i\Delta\omega\delta}\alpha\vec{U}_{CAV} .\tag{53}$$

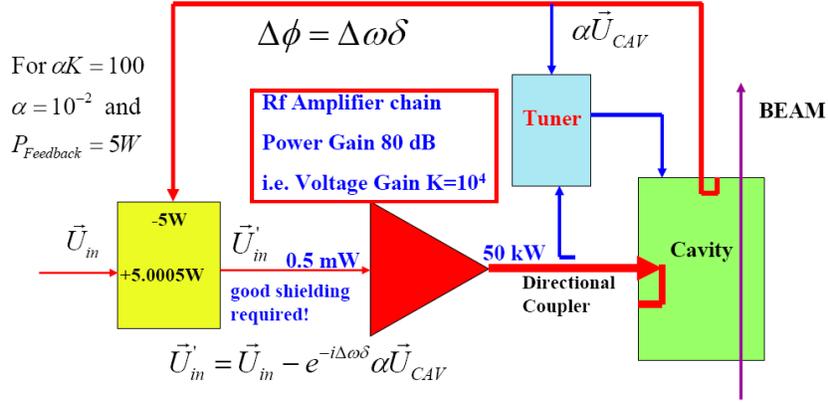

**Fig. 8:** Schematic of servo loops for phase and amplitude control of the HERA 208 MHz proton rf system

With the voltage gain $K$ of the amplifier we can rewrite Eq. (20) and obtain for the cavity voltage with feedback

$$\vec{U}_{CAV} \approx \frac{K\left[\vec{U}_{in} - e^{-i\Delta\omega\delta}\alpha\vec{U}_{CAV}\right] - \vec{U}_B}{1 + iQ_L 2\frac{\Delta\omega}{\omega_{CAV}}}\tag{54}$$

or

$$\vec{U}_{CAV} \approx \frac{K\vec{U}_{in} - \vec{U}_B}{1 + iQ_L 2\frac{\Delta\omega}{\omega_{CAV}} + e^{-i\Delta\omega\delta}\alpha K}\tag{55}$$

For $\Delta\omega = 0$ and $A_F \gg 1$ this reduces to

$$\vec{U}_{CAV} \approx \frac{\vec{U}_{in}}{\alpha} - \frac{\vec{U}_B}{\alpha K} .\tag{56}$$

The open loop feedback-gain $A_F$ is defined as

$$A_F = \alpha K .\tag{57}$$

One sees that there is a reduction of the beam-induced cavity voltage by the factor $1/A_F$ due to the feedback. This is equivalent to a similar reduction of the cavity shunt impedance as seen by the beam.

$$Z_L \approx \frac{R_{SL}}{1 + iQ_L 2\frac{\Delta\omega}{\omega_{CAV}}} \to \frac{R_{SL}}{1 + iQ_L 2\frac{\Delta\omega}{\omega_{CAV}} + A_F e^{-i\Delta\omega\delta}} .\tag{58}$$

The price for this fast reduction of beam-loading is the additional amount of generator current $\vec{I}_B N$ which is needed to almost compensate the beam current in the cavity. In terms of additional transmitter power $P'$ this reads

$$P' = R_S \hat{\vec{I}}_B^2 / 8. \tag{59}$$

It is the power already calculated by Eq. (45). Since there is no change in cavity voltage due to $P'$ this power will be reflected back to the generator, which has to have a sufficiently large plate dissipation power capability. Otherwise a circulator is needed. This critical situation of additional rf power consumption and reflection lasts, however, only until the tuner has reacted, and it may be minimized by pre-detuning. The generator-induced voltage is, of course, also reduced by the amount $1/A_F$, but this can easily be compensated on the low power level by increasing $\vec{U}_{in}$ by the factor $1/\alpha$ as Eq. (56) suggests. The practical implications of this will be illustrated by the following example.

Let the power gain of the amplifier be 80 dB. For a cavity power of 50 kW an input power $P_{in}$ of 0.5 mW is thus required. This corresponds to a voltage gain of $10^4$ so, for a design value of $A_F = 100$, $\alpha$ becomes $10^{-2}$. Hence the power, which is fed back to the amplifier input, is 5 W. In order to maintain the same cavity voltage as without feedback, $P_{in}$ has to be increased from 0.5 mW to 5.0005 W. This value can, of course, be reduced by decreasing $\alpha$. But then the amplifier gain has to be increased to keep $A_F$ constant. This leads to power levels in the 100 $\mu$W range at the amplifier input. All this is still practical, but some precautions, such as extremely good shielding and suppression of generator and cavity harmonics, have to be taken.

The maximum feedback-gain that can be obtained is limited by the aforementioned delay time $\delta$ of a signal propagating around the loop. According to Nyquist's criterion the system will start to oscillate if the phase shift between $\vec{U}_{in}$ and $\alpha \vec{U}_{CAV}$ exceeds $\approx 135°$. A cavity with high $Q$ can produce a ±90° phase shift already for very small $\Delta\omega$. Therefore, once the additional phase shift given by Eq. (52) has reached $\pm \pi/4$, the loop gain must have become $\leq 1$, i.e.,

$$\left| A_F(\Delta\omega_{max}) \right| \approx \frac{\alpha K}{1 + iQ_L 2\frac{\Delta\omega_{max}}{\omega_{CAV}}} \leq 1 \tag{60}$$

where

$$\delta \Delta\omega_{max} = \pm \frac{\pi}{4}. \tag{61}$$

Here we assume that all other frequency-dependent phase shifts, like the ones produced by the amplifiers, can be neglected. Inserting Eq. (61) in Eq. (60) we can solve for $A_F$:

$$A_F = \frac{Q_L}{4 f_{CAV} \delta}. \tag{62}$$

This is the maximum possible feedback-gain for a given $\delta$.

A fast feedback loop of gain 100 has been realized at the HERA 208 MHz proton rf system. With a loaded cavity $Q_L \approx 27000$ the maximum tolerable delay, including all amplifier stages and cables, is $\delta = 330$ ns. Therefore all rf amplifiers have been installed very close to the cavities in the HERA tunnel.

In addition, there are independent slow phase and amplitude regulation units for each cavity with still higher gain in the region of the synchrotron frequencies, i.e., below 300 Hz. Without fast feedback these units might become unstable at heavy beam-loading [8, 9] since changes then in cavity voltage and phase are correlated as shown by Eqs. (43) and (51).

The effect of a fast feedback loop is visible in Fig. 9 where the transient behaviour of the imaginary (upper curve) and real (medium curve) part of a HERA 208 MHz cavity voltage vector is displayed. The lower curve is the signal of a beam current monitor, which shows nicely the bunch structure of the beam and a 1.5 $\mu$s gap between batches of $6*10$ bunches each. A detailed description of this measurement and of the IQ detector used is given in Ref. [10]. In this particular case the upper curve is essentially equivalent to the phase change of the cavity voltage due to transient beam loading and the middle curve corresponds to the change in amplitude.

The apparent time shift between the bunch signals and the cavity signals is due to the time of flight of the protons between the location of the cavity and the beam monitor in HERA. The transients resulting from the first two or three bunches after the gap cause step-like transients, which accumulate without significant correction. Later the fast feedback delivers a correction signal, which causes the subsequent transients to look more and more saw-tooth-like. From this one can estimate the time delay in the feedback loop to be of the order of 250 ns. After about one 1 $\mu$s the equilibrium with beam is reached. Similarly, one observes in the left part of the picture that the feedback correction is still present during 250 ns after the last bunch, before the gap has left the cavity. The equilibrium without beam is also reached after about one 1 $\mu$s. Without fast feedback the time to reach the equilibrium is about 100 times larger, as one would expect for a feedback gain of 100.

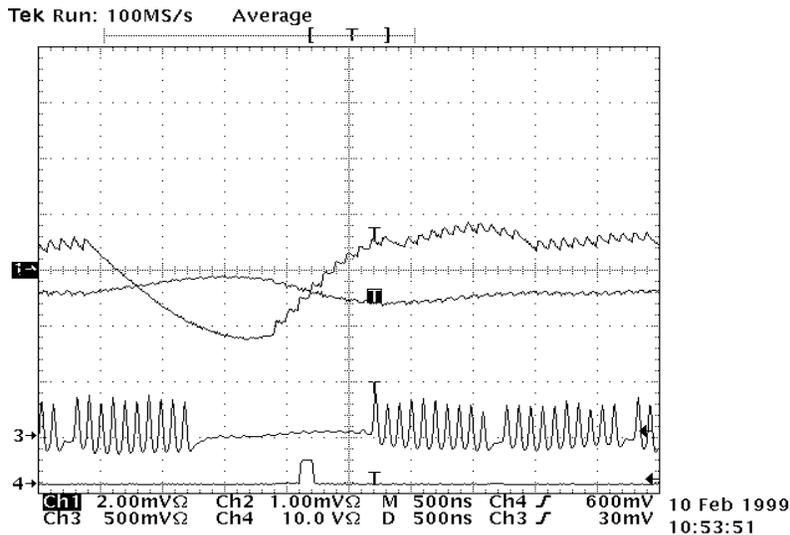

**Fig. 9:** Transient behaviour of the cavity voltage under the influence of fast feedback. This figure has been taken from Ref. [10].

To summarize this section we state that fast feedback reduces the resonant cavity impedance as seen by an external observer (usually the beam) by the factor $1/A_F$. It is important to realize that any noise originating from sources other than the generator, especially amplitude and phase noise from the amplifiers, will be reduced by the factor $1/A_F$ because the cavity signal is directly compared to the generator signal at the amplifier input stage. Care has to be taken that no noise be created, by diode limiters or other non-linear elements, in the path where the cavity signal is fed back to the amplifier input. This noise would be added to the cavity signal by the feedback circuit. This becomes especially important for digital feedback systems, where the digital hardware (downconverters, ADCs, DSP, etc.) is part of the feedback loop.

Amongst the great advantages of digital technology are very easy amplitude and phase control of each channel (analog elements are very expensive), easy application of calibration procedures and factors etc. However, there can be the disadvantage of very high complexity.

## 5. Feedback and feed-forward applied to superconducting cavities

So far, we have mainly considered normal conducting cavities in a proton storage ring, where the protons arrive in the cavities at the zero crossing of the rf signal, i.e., at $\phi_s = 0°$ or a few degrees. In the following we would like to present an example of the other extreme: superconducting cavities in a linear electron accelerator where the electrons cross the cavities near the moment of maximum rf voltage, i.e., at $\phi_s \approx 90°$. (Note that for linear colliders a different definition of $\phi_s$ is usually used, namely $\phi_s = 0°$ when the particle is on crest. In this article we do not adopt this definition.)

In the beginning of the 1990s a test facility for a TeV Energy Superconducting Linear Accelerator (TESLA) was erected at DESY. In the meantime a unique world-wide Free Electron Laser user facility named FLASH, which is generating photon beams in the nm wavelength range for a rapidly growing user community, has emerged from this test facility. We refer to the special example of the superconducting nine-cell cavities of this accelerator, which are made of pure niobium. The operating frequency is 1.3 GHz.

The unloaded $Q_0$ value of these cavities is in the range $10^9 - 10^{10}$, or even higher. Hence the bandwidth is only of the order of 1 Hz, and also the superconducting cavity shunt impedance exceeds that of normal conducting ones by many orders of magnitude. Since the particles are (almost) on crest, only the real part of the cavity admittance as seen by the generator Eq. (27) is changed due to beam loading. This means that for beam loading compensation only a change in the coupling factor $\beta$ is required and detuning plays no role for beam loading compensation in this situation. There is only perfect matching for the nominal beam current to which the cavity power input coupler has been adjusted. As we have already mentioned in Section 2.2, it takes the value $\beta = 6401$ in this case, which reflects also the fact that the ratio of the power taken away by the beam to the power dissipated in the cavity walls is much larger for superconducting cavities than for normal conducting ones. Because of the coupling, the nominal loaded $Q_L$ value is only $3*10^6$, and the corresponding cavity bandwidth is 433 Hz. Since in this case there is a circulator with a load to protect the klystron from reflected power, the rf generator always sees a matched load.

From the circuit diagram in Fig. 10 we see that one rf generator supplies up to 32 cavities with rf power. The rf power per cavity needed to accelerate an electron beam of 8 mA to 25 MeV amounts to 200 kW, hence the minimum klystron power needed is 6.4 MW. This power is entirely carried away by the beam. In contrast to the previous example, where all the rf power was essentially dissipated in the normal conducting cavity walls, the power needed to build up the rf cavity voltage in the superconducting cavities is only a few hundred watts. Additional rf power is needed to account for regulation reserve, impedance mismatches, etc. Therefore high-efficiency 10 MW multi-beam klystrons were developed for this project. For completeness we mention that this is pulsed power, with a pulse length of 1.5 ms and the maximum repetition rate 10 Hz. So the maximum average klystron power is 150 kW.

The rf seen by the beam corresponds to the vector sum of all cavity signals. Therefore, in a first step, this vector sum must be reconstructed by the low-level rf system. This is done by down-conversion of the cavity field probe signals to 250 kHz intermediate frequency signals, which are sampled in time steps of 1 $\mu$s. Each set of two subsequent samples corresponds then to the real and imaginary part of the cavity voltage vector. From these signals the vector sum is generated in a computer and is compared to a

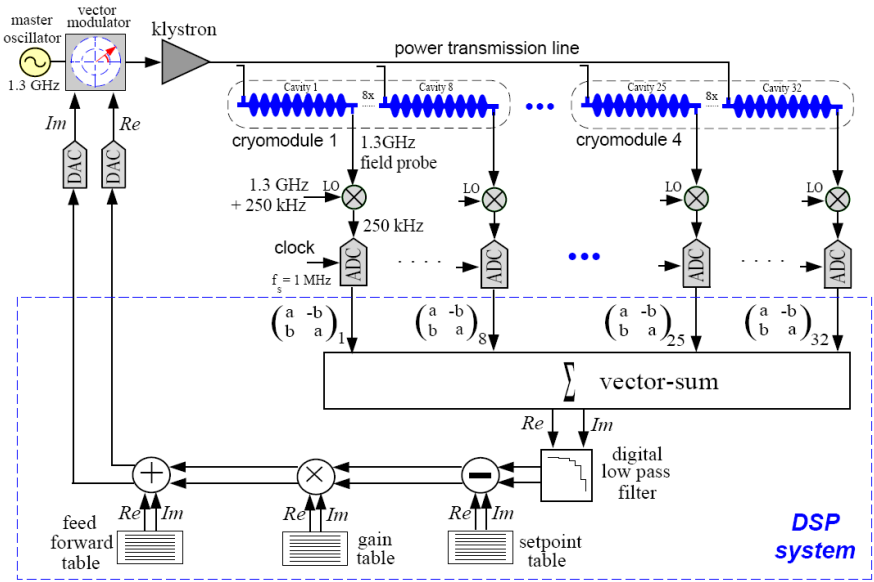

**Fig. 10:** Schematic of the low-level rf system for control of the rf voltage of the 1.3 GHz cavities in the TESLA Test Facility. This figure has been taken from Ref. [11].

table of set point values. The difference signal, which corresponds to the cavity voltage error, acts on a vector modulator at the low-level klystron input signal. In addition to this feedback a feedforward correction can be added. The advantage of feedforward is that, in principle, there is no gain limitation as in the case of feedback. If the error is known in advance, one can program a counteraction in the feedforward table. Examples for such errors could be a systematic decrease of beam current during the pulse due to some property of the electron source, or a systematic change of the cavity resonance frequency during the pulse. This effect exists indeed. The mechanical forces resulting from the strong pulsed rf field in the superconducting cavities cause a detuning of the order of a few hundred Hz at 25 MV/m. This effect is called Lorentz force detuning.

From Eq. (62) one might infer that due to the large value of $Q_L = 3*10^6$ the maximum possible feedback gain in this case could become significantly larger than for normal conducting cavities. However, one has to check whether there are poles in the system at other frequencies, and, at least in this case, there is a fairly large loop delay of about 4 $\mu$s caused by the 12 m length of the cryogenic modules in which the cavities are placed and by the time delay in the computer. This results in a realistic maximum loop gain of 140.

The most impressive results for amplitude and phase stability recently obtained with the newly installed third harmonic rf system of the FLASH accelerator [12] are shown in Figs. 11–14. The digital rf control system used here has the same basic structure as that indicated in Fig. 9. But, in addition, there is a digital MIMO (Multiple in Multiple out) controller in the feeedback path and also a learning feedforward system, which is described in detail in Ref. [13].

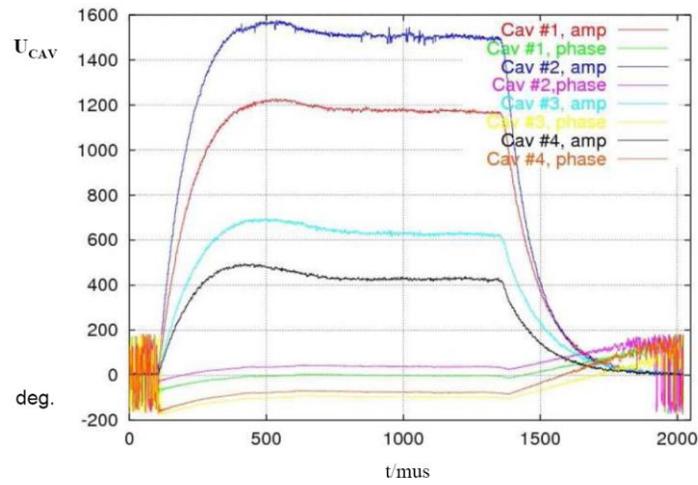

**Fig. 11:** Unregulated signals of rf phase (the lower four curves) and amplitude of four superconducting cavities operating at 3.9 GHz in the FLASH accelerator [14]

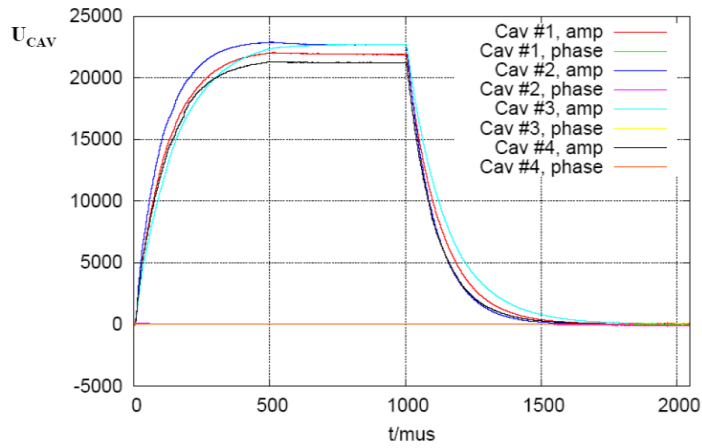

**Fig. 12:** Regulated signals of rf amplitude of four superconducting cavities operating at 3.9 GHz in the FLASH accelerator [14]

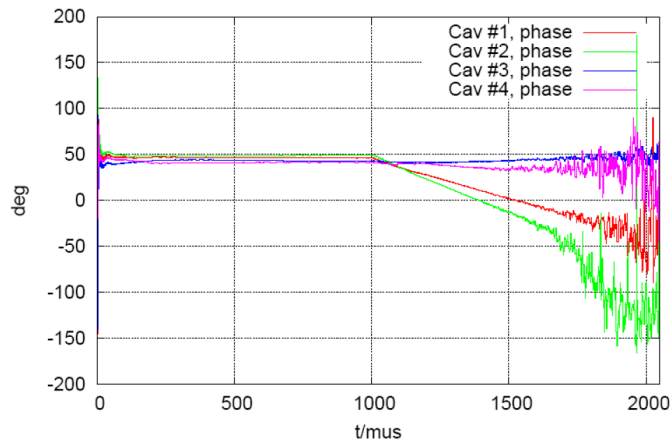

**Fig. 13:** Regulated signals of rf phase of four superconducting cavities operating at 3.9 GHz in the FLASH accelerator [14]

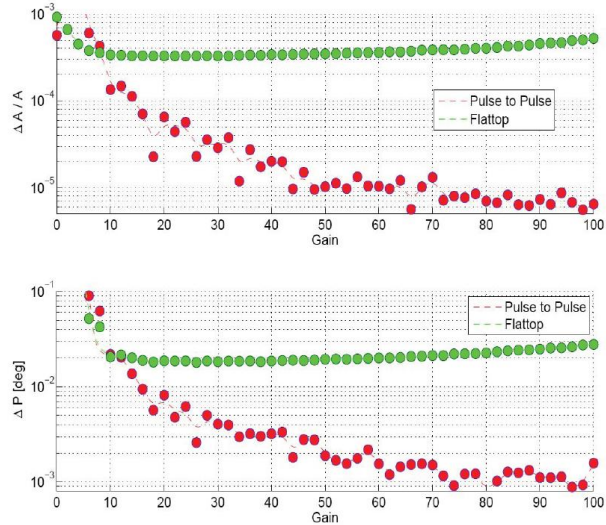

**Fig. 14:** Phase and amplitude rms stability vs. feedback gain achieved by digital feedback with integrated MIMO controller and learning feedforward in 3.9 GHz cavities operating in the FLASH accelerator as a 3$^{rd}$ harmonic system. The improvement in the pulse to pulse results is due to the effect of averaging over the measurement noise. This figure has been provided courtesy of C. Schmidt [15].

## 6. Damping of synchrotron oscillations of protons in the PETRA II machine

In the preceding sections phase and amplitude control of the cavity voltage was discussed. In this last section we would like to give an example of beam control by means of a dedicated rf system for damping synchrotron oscillations of protons in the PETRA II synchrotron at DESY.

Prior to injection into HERA, protons were pre-accelerated to 7.5 GeV/*c* and 40 GeV/*c* in the synchrotrons DESY III and PETRA II, respectively [16]. Timing imperfections during transfer of protons from one machine to the next one and rf noise during ramping were observed to cause synchrotron oscillations which, if not damped properly, may lead to an increase of beam emittance and to significant beam losses. Therefore a phase loop acting on the rf phase to damp these oscillations of the proton bunches was a necessary component of the low-level rf system. The PETRA II proton rf system, which consisted of two 52 MHz cavities, each with a closely coupled rf amplifier chain and a fast feedback loop of gain 50, was similar to the one shown in Fig. 8. The block diagram of the PETRA II phase loop, on which we will concentrate now, is shown in Fig. 15.

### 6.1 Loop bandwidth

The maximum number of bunches was 11 in DESY III and 80 in PETRA II so that eight DESY III cycles were needed to fill PETRA II. If synchrotron oscillations due to injection timing errors arise, all bunches of the corresponding batch are expected to oscillate coherently. Therefore one single correction signal can damp the bunch oscillations in that batch and in total up to eight such signals were needed, one for each batch. This phase loop was a batch-to-batch rather than a bunch-to-bunch feedback. Ideally, the correction of expected errors of about two degrees in the injection phase had to be switched within the 96 ns separating the last bunch of batch n from the first one of batch n + 1. Owing to the fast feedback of gain 50 the rf system had an effective bandwidth of about 1 MHz; it was, however, capable of performing small phase changes of the order of 1° per 100 ns, which was sufficient for damping synchrotron oscillations also in the multi-batch mode of operation.

## 6.2 The phase detector

Each bunch passage generates a signal in the inductive beam monitor also shown in Fig. 15. A passive LC filter of 8 MHz bandwidth filters out the 52 MHz component. The ringing time is comparable to the bunch spacing time as shown in Fig. 16. Amplitude fluctuations of this signal are reduced to ±0.5 dB in a limiter of 40 dB dynamic range. So the amplitude dependence of the synchrotron phase measurement between the bunch signal and the 52 MHz rf source signal is minimized. The phase detector has a sensitivity of 10 mV per degree. By inserting a low pass filter one can directly observe the synchrotron motion of the bunches at the phase detector output. This is shown in Fig. 18(a) for one batch of nine proton bunches circulating in PETRA II with the momentum of 7.5 GeV/$c$ a few ms after injection. The observed synchrotron period $T_S$ = 5 ms agrees with the expected value for the actual rf voltage of 50 kV.

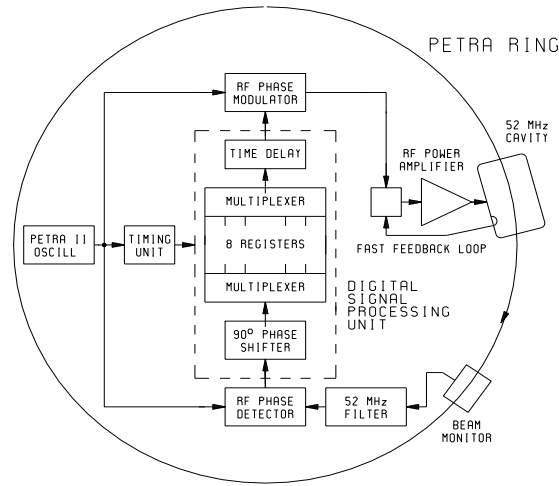

**Fig. 15:** Block diagram of the PETRA II phase loop. In the phase detector, synchrotron oscillations of the bunches are detected by comparing the filtered 52 MHz component of the beam to the 52 MHz rf reference source. An average phase signal for each of the eight batches of ten bunches is phase shifted by 90° with respect to the synchrotron frequency, stored in its register, and properly multiplexed to the phase modulator acting on the rf drive signal.

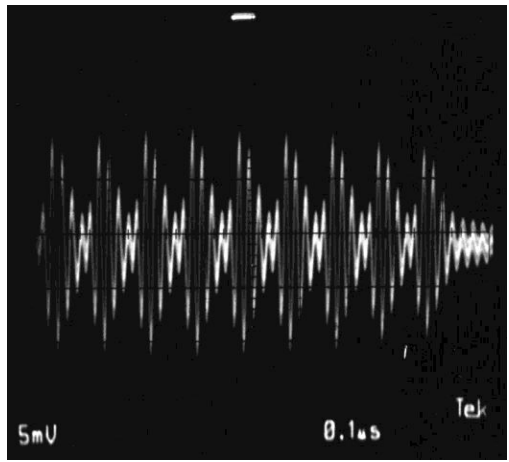

**Fig. 16:** Filtered signal of a batch of nine proton bunches circulating in PETRA. The bunch spacing time is 96 ns.

## 6.3 The FIR filter as a digital phase shifter

A feedback loop can damp the synchrotron motion if, as indicated in Fig. 15, the synchrotron phase signal is shifted by −90° relative to the synchrotron frequency $f_S$, delayed properly, and fed into a phase modulator acting on the 52 MHz drive signal. The necessity of the −90° phase shift relative to $f_S$ can be seen from the equation of damped harmonic motion $\ddot{x} + a\dot{x} + bx = 0$ with the solution $x = A\sin(\omega_S t - \phi)e^{-at}$. The damping term $a\dot{x}$ is proportional to the time derivative of the solution $x$, i.e., a phase shift of −90°. The correction signal will coincide with the corresponding batch in the cavity if the total delay $\tau = t_f + nT_{rev}$, where $t_f$ is the transit time from the beam monitor to the cavity, $n$ an integer, and $T_{rev}$ = 7.7 µs is the particle revolution time in PETRA. Sinc $T_S \gg T_{rev}$, a delay of even more than one turn ($n > 1$) would not be critical.

Rather than using a simple RC integrator of differentiator network as a −90° phase shifter, which is not without problems [17], a more complex digital solution with a software controlled phase shift has been adopted. This is very attractive since during injection, acceleration, and compression of the bunches the synchrotron frequency varies in the range from 200 Hz to 350 Hz. In addition, storing and multiplexing the eight correction signals for each of the eight possible batches in PETRA II can also be realized most comfortably on the digital side. The phase shifter has been built up as a three-coefficient digital FIR (Finite-length Impulse Response) filter according to

$$g_\mu = \sum_{k=0}^{2} h_k f_{\mu-k},\qquad (63)$$

with an amplitude response

$$H(\omega) = \sum_{k=0}^{2} h_k e^{-ik\omega T_S} \qquad (64)$$

where $f$ and $g$ are input and output data respectively. Using the coefficients $h_0 = \frac{2}{\pi}\sin\phi, h_1 = \cos\phi, h_2 = -\frac{2}{\pi}\sin\phi$ one obtains a phase shift which, in the frequency range of interest $200\,\text{Hz} \leq f_S \leq 359\,\text{Hz}$, deviates by less than ±0.4 from the nominal value $\phi = -\frac{\pi}{2}$ in accordance with Eqs. (63) and (64). The frequency dependence of the phase shift is mainly due to the delay in the filter which is of the order of 1 ms, i.e., two sampling periods. It can always be corrected by software, if necessary. The amplitude response is constant within a few per cent for all frequencies.

A block diagram of the filter is shown in Fig. 17. The synchrotron phase information of the eight batches is sampled at intervals $T_S$ = 0.5 ms and passed through eight times three shift registers. The three coefficients are stored in ROMs and are appropriately combined with the phase information. So, the first filter output is available after three sampling periods and is then renewed every 0.5 ms.

The performance of the loop is demonstrated in Fig. 18 where the phase detector output recorded by a storage scope is displayed. Complete damping of the synchrotron oscillation is achieved within less than one period. This corresponds to a damping time of less than 4 ms. If the loop is operated in the anti-damping mode, the beam is lost within some ms. With the loop, losses of the proton beam in PETRA II during energy ramping could be significantly reduced.

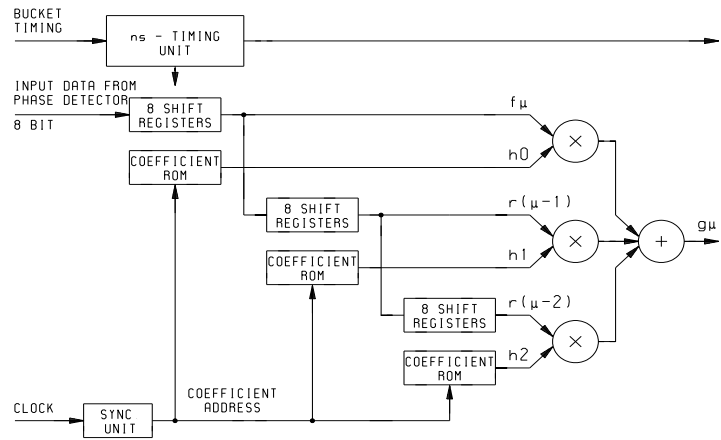

**Fig. 17:** Block diagram of the FIR filter. From three successive sampling periods the averaged phase signals for the eight proton batches in PETRA II are stored in shift registers and combined with the three coefficients, which are stored in ROMs. The first phase-shifted output is available after three sampling periods of 0.5 ms and is renewed every sampling period.

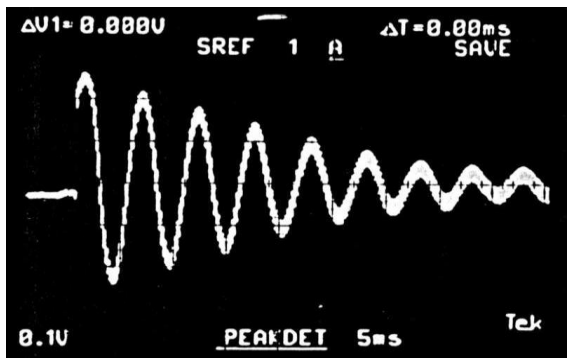

**Fig. 18(a):** The synchrotron oscillation measured at the phase detector output a few ms after injection of a batch of nine proton bunches into PETRA II. It is smeared out by Landau damping after some periods. The damping loop is not active.

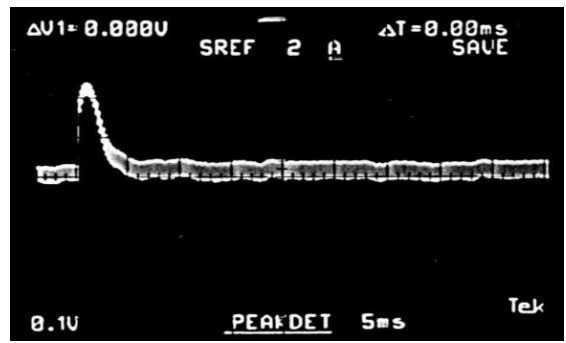

**Fig. 18(b):** Same as Fig. 18(a) but with the phase loop active. The synchrotron oscillation is completely damped within half a synchrotron period of 0.5 ms.